\documentclass[10pt,conference]{IEEEtran}

\usepackage{graphicx}
\usepackage{comment}
\usepackage{booktabs}
\usepackage{float}
\usepackage{adjustbox}
\usepackage{lscape}
 \usepackage{multirow}
 \usepackage[normalem]{ulem}
 \useunder{\uline}{\ul}{}
 \usepackage{lscape}
 \usepackage{longtable}
 \usepackage{array}
 \usepackage{url}
 
\title{Exploring Context-Aware Conversational Agents in Software Development}

\author{\IEEEauthorblockN{Glaucia Melo, Edith Law, Paulo Alencar, Don Cowan} 
\IEEEauthorblockA{\textit{David R. Cheriton School of Computer Science} \\
\textit{University of Waterloo}\\
Waterloo, Canada \\
\{gmelo,edith.law,palencar,dcowan\}@uwaterloo.ca}}

\begin{document}

\maketitle

\begin{abstract}
Software development is a complex endeavor that depends on a wide variety of contextual factors involving a large amount of distributed information. This knowledge could include: technology-related tasks, software operating environments and stakeholder requirements. A major roadblock to using this knowledge in software development is that most of this information is implicit and captured in the developers’ minds (tacit) or spread through volumes of documentation. Developers, as they work often have to maintain mental models of these tasks as they produce the software. As a result, context can be easily lost or forgotten and developers often use trial-and-error approaches while finishing the project. This study aims at analyzing whether supporting software developers with a chatbot during task execution can improve the overall development experience. The chatbot can assist the  developers with executing different tasks based on implicit contextual information. We propose an implementation to explore the viability of using textual chatbots to assist developers automatically and proactively with software development project activities that recur.

\end{abstract}


\section{Introduction}

Software development is a complex activity, with an increasing demand for professionals. Developers perform various software development tasks, with different context switches \cite{bradley2018context, meyer2017work}. Most of this context, which can include: tool preferences, approaches to tasks, documentation, and the project, is tacit, and brings significant  added complexity to software development. Developers maintain mental models of a project combining items such as various types of data including system, version control, environment, project details and tasks \cite{bradley2018context}, and it would be beneficial if at least some of these details did not have to rely on a developers’ ability to remember the execution sequence. A simple solution to support developers when executing tasks, they must remember should be to automate these tasks. However, automation leads to a set of challenges, as indicated by Bradley et al. \cite{bradley2018context} because:

\begin{itemize}
\item It is hard to determine in advance all the tasks and workﬂows developers will need to complete;
\item These workflows consist of various actions that often span tool boundaries and require a diverse set of parameters that may depend on the current context and developer intent;
\item Even if there are scripts configured for automating workflows, the developer needs to remember their existence and how to invoke them in the current context.
\end{itemize}

There are also other examples of activities that are hard to automate, such as searching for support in an implementation, or editing a set of three files but wanting to commit only two, because the developer thinks the third one needs further review. Moreover, in software development, having the human-in-the-loop has proven to be beneficial in terms of searching for support when developing project tasks  \cite{melo2019retrieving}. 

A few examples of project tasks are: the next step in process (open task / code / test / deploy / close task), some recurrent commands (git push/pull, deploy, close task/issue), the available testing options, and code review options, determining the developers’ next assigned task, deploying workflow based on the environment and is there support available for this task (API Tutorial, Stack Overflow post, commit commented).
In this paper, we conduct an exploratory study that aims at understanding the variety of tasks in software development that could be supported by a textual conversational agent (chatbot). We want to identify through a pilot study how developers interact with the agent, and developers’ preferences if the agent can support the execution of current project tasks and not just be a developer’s memory (mind maps).

We aim to investigate this problem of developer support through the following research question: \textbf{How well can a chatbot support software developers while executing development tasks?} We believe that the use of a proactive chatbot can make these mostly implicit contexts more explicit and make the development process less prone to errors as the developer does not have to remember all the tasks to be performed.
We hypothesize that capturing context (embedded in software development tasks) in software development and having a chatbot to suggest these tasks to developers can leverage software development, as recurrent and implicit tasks can be forgotten. For this exploratory pilot study, we performed a Wizard-of-Oz study 
using Slack as the platform to support developers. We also interviewed participants (using a questionnaire) at the beginning and end of the experiment. Preliminary results indicate that developers are interested in using chatbots as a support tool during software development, with a focus on using  the chatbot for task management purposes.

We plan to contribute through performing what is, to the best of our knowledge, the first experiment that aims at capturing possible preferences of software developers when interacting with chatbots. Additionally, we identify as a contribution, the empirical mixed-method study (Wizard-of-Oz + questionnaire) to evaluate the questions software developers would ask when interacting with chatbots in their daily work.

The remainder of this paper is organized as follows: Section \ref{relatedwork} presents important related literature. Section \ref{illustrative} presents an example of a situation where the system could be useful and how it might operate. Sections \ref{studydesign} and \ref{pilot1results} present the study design and results, respectively. Lastly, Section \ref{conclusion} concludes the paper. 

\section{Related Work} \label{relatedwork}  

Researchers have studied software development contexts with a view to supporting developers in their tasks. Automating development tasks has also been presented as a means to improve software development, including the use of bots in Software Engineering (SE). Another emerging solution is the use of machine learning to manage dialogue between humans and bots. In this context, we present Rasa, an open-source tool that uses interactive learning and other features for building conversational software, as an example of a tool that could support the evolution of this study.

\subsection{Context in Software Development}

Research has defined context in computing as the information that is part of an application’s operating environment and that can be sensed by the application  \cite{salber1999context}. The same work mentions that context-aware applications, which sense context information and modify their behavior accordingly without explicit user intervention, are of major interest.

A broad corpus of research presents ways in which context in software development might ease the work of a human trying to perform a task. As stated by Murphy et al., leveraging context using smart assistants, for example, could allow developers to remain focused on hard parts of problems while the assistant keeps tracks of normal tasks \cite{murphy2019beyond}. Although latent, the context in software engineering can also be implicit. Therefore, some studies might have worked with the context in SE but without explicitly defining context nor having its usage motivated \cite {gasparic2017context}. Nonetheless, several studies have considered the context in software development and identified it as "context" \cite{melo2019context}. 

Gasparic et. al. \cite {gasparic2017context} propose a context model that aims at supporting and enhancing several essential applications related to developers’ interactions with IDEs. In this paper, we propose the use of a software development context that is not restrained to the current IDE, but supports different and IDE-independent development tasks. These tasks could be happening in different stages of the development process including coding, testing or deployment.

Other research has used context in the sense of the support of software  producers’ needs while developing software \cite{melo2019retrieving}. This approach is interesting as it concludes that this specific context information such as knowledge from Stack Overflow, can be reused if documented. With the proposal in this paper, the context can also be a source of support for developers, indicated/recommended by the chatbot either proactively or while sensing that the developer needs help. This information such as the support needed as a Stack Overflow post, may or may not be available depending on the current context of the developer.

Meyer et al. \cite{meyer2017work} have considered context in the sense of how switching between contexts while developing can interfere in a software developer’s productivity. This work is really interesting because it contributes to how developers perceive their productivity, and more importantly, how switching between tasks gives developers a sense of diminishing their productivity. In this paper, we suggest developers’ tasks that should be done for a specific context, guiding them to the conclusion of a more significant goal. An approach that could guide developers back to what they have to do in case their context suddenly changes.

In all, we aim to guide and support the human who is in-the-loop of a software development process to execute tasks that can differ when the context is different. Bringing humans into the loop might result in a bot that is smarter than the ones reported in recent work \cite{marieli2018}. The integration of bots and humans brings a sense of partnership, where one can complement the other, while integrating different contexts in one single interaction channel.

\subsection{Conversational Agents in Software Development}

Conversational agents (CAs) have been around since the 1960s when Joseph Weizenbaum developed ELIZA to reproduce typical human interactions between a user and a computer \cite{weizenbaum1966eliza}. CAs development assumes people interacting with systems that are able to capture, interpret and respond in natural language, comparable to engaging in a conversation with another human being \cite{mctear2016rise}. 

In software development, bots are used to support different activities, from automating repetitive tasks to bridging knowledge and communication gaps in software teams \cite{storey2016disrupting}.
Our proposal sits within the scope of providing means for developers to interact with a system that is supposed to support them during development, through a conversation where the chatbot should be aware of what is there for the developer to do (supported by a context-model). This means the chatbot should know both the workflow and the contextual information such as the repository, projects and commands of the developer.

Bradley and colleagues \cite{bradley2018context} have considered a context-model for supporting conversational developer assistants that utilize the context elements needed to support workflow involving a distributed version control system that starts to support the assistant concept just outlined. This design is very similar to the proposal in this paper. The difference comes when those authors investigate voice control agents for version control tasks. Moreover, our intention is guiding developers in what they have to do more than executing the task for them. We believe that the nature of capturing the different contexts and presenting to developers the tasks related to that context is already a significant contribution, similar to what Bradley and colleagues have performed.

Other work has investigated two different aspects of the usage of chatbots in the realm of software engineering: to detect code conflicts when developers are working in parallel \cite{paikari2019chatbot} and as an expert recommendation system to help developers find the right person to contact within open source projects \cite{cerezo2019building}. Both methods are very interesting, but differ from the approach in this paper. We aim to identify the developers’ current context, guiding them through the tasks that are expected to be executed in that specific context with the help of a proactive chatbot assistant.

\subsection{Chatbots and Rasa}

CAs can be classified by their mode of communicating (voice-activated, text-based or embodied) \cite{gnewuch2017towards} \cite{lee2009example}, and also whether their setting is general-purpose or task-oriented. Other kinds of CAs consider usability, efficiency, effectiveness and user satisfaction \cite{radziwill2017evaluating}. A chatbot is a software application used to conduct an online chat conversation via text or text-to-speech, in lieu of providing direct contact with a live human agent\footnote{\url{wikipedia.org/wiki/Chatbot}}.

Several different chatbot tools have been developed to promote conversational support including: Botpress\footnote{\url{botpress.com}}, Google Dialogflow\footnote{\url{dialogflow.com}}, Microsoft Bot Framework\footnote{\url{dev.botframework.com}}, Wit.ai\footnote{\url{wit.ai}} and Amazon Lex\footnote{\url{aws.amazon.com/lex}}. These tools serve a similar purpose, which is to support the development of services for conversational interfaces using voice or text. Modern platforms include artificial intelligence (AI) approaches to automate communication and create scalable personalized customer experiences \cite{brenier_prial_2019}. These solutions, although serving a similar purpose, might have specific properties. For future work, we intend to use a tool that can bootstrap its operation from minimal (or no) initial training data, and that also can manage different contexts in its conversation. Being open-source and easy to use, as well as having decoupled features, are also reasons why we chose Rasa\footnote{\url{rasa.com}} \cite{bocklisch2017rasa} for future work. Rasa is an open-source framework. It has two major components: Rasa NLU and Rasa Core. Rasa NLU is responsible for natural language understanding. Rasa Core is a framework for building a conversational chatbot. Rasa core allows sophisticated dialogue, which is trained using interactive and supervised machine learning. The key benefit of using the open-source NLU module is that you don’t need to send your data outside to Google or Amazon to train on intents and entities. Being open source, you can tweak the models or develop models to satisfy your needs. The architecture for the Rasa NLU also allows it to run from anywhere as a service \cite{singh2019introduction}.

\section{Illustrative Example and Potential System} \label{illustrative}

Software development is a complex endeavor that usually involves a team of people with different backgrounds and expertise, a variety of tools and different domains. Developers have an end-goal, and this goal can be broken into smaller sub-goals to be managed and achieved by the members of the team. Of course, these sub-goals also have to comply with existing workflows.

Consider the example illustrated in \cite{melo2019context}, where Gabi is a software developer who has been programming in Java for nine years and has recently been working on project X. When there is a new project, one of Gabi’s goals is to create a minimal viable product (MVP) to show her clients. She deploys the software locally, using a container tool such as Docker and manually uploads the project to a web server. Gabi also reboots the server manually after each deployment, so changes become effective. This approach is faster, and she does not have to configure a job or a server to generate a deployment automatically, which would cause a longer wait for her clients to see the MVP.

When a version of the software in project Y, a huge mature project, has to go to production, all Gabi does is commit the code from the local to the shared code repository. Then, the scheduled automated job in a Jenkins server will take care of the other steps, which are checking out the code, building the project, uploading the package on the server and rebooting the server.

\begin{figure}[]
    \centering
    \includegraphics[scale=0.5]{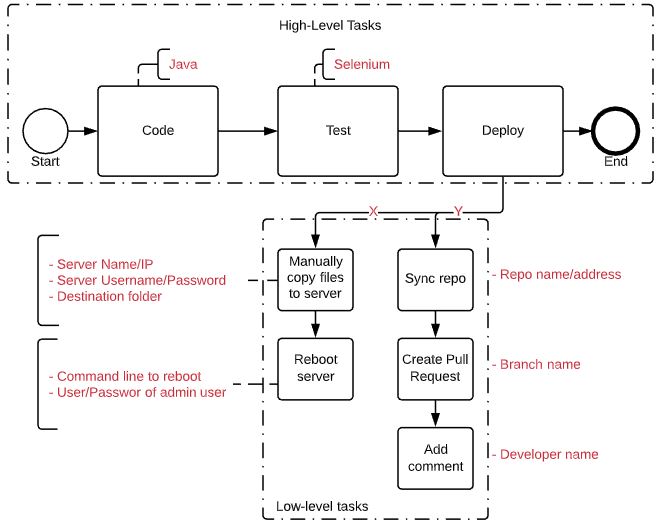}
    \caption{Example of Workflow with High and Low-Level Tasks.}
    \label{workflow}
\end{figure}

In theory, the steps for Project X and Y are the same (deploy), but because the projects are different, Gabi’s work is different. Since the context, namely the  Project differs, the context defines the steps for the deployment phase. If Gabi, who has been working on project Y for years, forgets she needs to deploy Project X manually, this can create a problem. A representation of this situation and different possible contexts are shown in Figure \ref{workflow}. The information in red represents the context that can vary in this example, and therefore, influence the workflow.

This figure does not represent several other contextual attributes, including team expertise, hardware and software technologies, IDE tools, as well as location and time-zone. These other attributes could also influence Gabi’s workflow.

The idea is to provide a chatbot that recognizes the developer’s current context and is able to suggest the steps the developer should follow. The workflow and context should be embedded in the conversational agent’s understanding, and tracked by the chatbot. The following conversation, for example, could occur between Gabi and the chatbot.

\textbf{Chatbot: }Hi! \\
\textbf{Gabi: }Hi, I need to \textit{deploy} project \textit{x} in \textit{development} environment.\\
\textbf{Chatbot:} Sure. You should manually copy the files to server 10.1.1.1, into folder \url{\\usr\\system\\folder1}. Use your username gabidev to access the server. \\
\textbf{Gabi:} Alright, I copied the files. What is next? \\
\textbf{Chatbot:} You should reboot the server using this command: pseudo-reboot-n. The password is 1234567890. \\
\textbf{Gabi:} What's next? \\
\textbf{Chatbot:} You are all done! Please check the application using the following link \url{localhost:8000/mvp}. 

Knowing all these different contexts are available and could be communicated to developers using a chatbot, we then execute a study to understand the preferences of software developers when using chatbots to support their work. The next section presents the study, followed by the results.

\section{Pilot Study Design} \label{studydesign}

The goal of our experiment is to understand how software developers would use a chatbot to facilitate daily software engineering tasks and processes. For that, we aim to provide a chatbot interface to developers, and using an example scenario of what developers might perform in one day and the related questions. We would then capture these questions and analyze the ones that were asked. In effect, we want to investigate that if developers are given a chatbot they are willing to use for their daily work, for what kinds of tasks would this chatbot be useful. Further, are there unexpected questions/requests that were not anticipated. In this context, we examine the following research question:

\textbf{RQ:} How well can a chatbot support software developers while executing development tasks?

To attempt an answer to this research question, we have conducted an experiment with graduate students as participants. First, we give participants a scenario of a day in the life of a software developer. Then, we ask participants to ask questions regarding any step of this scenario. We intend to determine if there is one part of the scenario participants ask more or fewer questions about (interest) and the types of words and questions used (vocabulary).

\subsection{Procedure Setup}

We use a mixed-method experiment, combining Wizard-of-Oz (WOZ) \cite{woz} \cite{kelleydid} and a structured questionnaire. We decided to do the pilot study using WOZ to test if there would be user engagement and understanding of the study before we actually invested time in developing a real chatbot. The experiment is organized in three different steps.

\textbf{Step 1:} a questionnaire to collect information from the participants and explain the procedures of the study and hand them the scenario, which participants can read. Participants can keep the scenario for reference and read it anytime during the study.

\textbf{Step 2:} an experiment with the WOZ using Slack, where participants after being presented with a scenario, are supposed to type the questions they would ask on the chatbot interface.

\textbf{Step 3:} a post-study questionnaire to gather participant’s opinions regarding the usefulness of the prototype, completing the Likert scale and gather suggestions and impressions from participants.

\subsection{Participants}

For this study, students from the Conversational Agents Winter 2020 course (David R. Cheriton School of Computer Science, University of Waterloo) were recruited. For step 1, we have collected the following information from the participants using Google Forms:

\begin{enumerate}
\item Current degree: Undergrad, Masters or Ph.D. Student
\item Background (Program and area attending)
\item Experience with software development (in months)
\item Knowledge regarding software repositories (Git/Jira)
\end{enumerate}

After completing the demographic data, the participants were presented with a scenario, as step 1 of the study. The scenario, which is presented next, was created after asking seven software developers currently working in the industry about their workday activities. They were asked to describe a day in their office, and we have merged the answers to create a feasible, realistic, coherent scenario.

\textit{\textbf{Scenario}: You arrive at your office at 8 am, sit on your chair and open the task manager to check what tasks are assigned to you. You look for the one task with the highest priority and read the text of the task. You understand what you are supposed to do for this task. You have a stand-up meeting and redefine task priority. You go to the cafeteria and grab a coffee. You’re back to your desk and open Eclipse. You sync your local code repository to get the newest version of code for that specific task. You look for the artifacts that you might have to edit to complete this task. You ask your colleague what he thinks about the task and the artifacts you decided to edit. You start coding the test case for that new implementation and then edit the .java class that is supposed to receive the edits. You get another coffee. You save your edits, commit your code and create a pull request, inserting a comment in the pull request. It’s 4 pm. You update the task assigned to you to status Done and with the time you invested in solving the task. You leave the office.}

After reading and understanding the given scenario, participants were supposed to interact with the chatbot, by asking any question they thought to be adequate. The chatbot is prepared to answer regarding the possible questions related to the scenario above and deal with greetings and exceptions. All cases with prepared answers are listed in Table \ref{tab:answers}. 

\begin{table*}[]
\centering
\caption{Pre-configured answers for Study.}
\label{tab:answers}
\begin{tabular}{|l|l|}
\hline
\multicolumn{2}{|l|}{\textbf{Greetings}} \\ \hline
Greet & Hey! What do you need to know regarding your scenario today? \\ \hline
Goodbye & Bye!! \\ \hline
\multicolumn{2}{|l|}{\textbf{Scenario}} \\ \hline
Artifacts & You need to edit class client.java and class diagram OrderMain. \\ \hline
Commit & OK, I am committing your code to your repository. \\ \hline
Tasks & Tasks 27405, 67933 and 13003 are open and assigned to you. \\ \hline
Priority & The task with the highest priority assigned to you is Task 13003. \\ \hline
Pull Request & Please, create a pull request to a branch identified with your task ID. \\ \hline
Code Push & \begin{tabular}[c]{@{}l@{}}Please, download or clone your repository to your local machine. \\ Keep it synchronized.\end{tabular} \\ \hline
Sync & \begin{tabular}[c]{@{}l@{}}You should synchronize your repo. Click in the sync repo \\ button on your version control page.\end{tabular} \\ \hline
Tool & \begin{tabular}[c]{@{}l@{}}You have to open Eclipse or VSCode and synchronize your code. \\ After, assign a test task to the test team.\end{tabular} \\ \hline
\multicolumn{2}{|l|}{\textbf{Exceptions}} \\ \hline
No answer & Did that help you? What else can I help you with? \\ \hline
What can the bot do & \begin{tabular}[c]{@{}l@{}}I can help you discover which tasks are assigned to you, \\ their priorities, how to do a pull request and you can ask me \\ to commit your code.\end{tabular} \\ \hline
Not in the list & Sorry, I can't help you with that. Is there anything else I can help with? \\ \hline
\end{tabular}
\end{table*}

Then we ask the participants:  What questions would you be willing to ask your assistant bot? Participants are then invited to talk to "DevBot" on Slack. Because it was a WOZ study, DevBot was actually one of the authors of this paper, answering participants according to the preconfigured answers in Table \ref{tab:answers}\footnote{"What can the bot do" answer was presented either when the participant asked explicitly what the bot could do, or after the third time the bot had to fallback.}. When participants asked for the bot to execute tasks repetitively, the chatbot responds with "Done." and asks the question configured in "No answer", to stimulate the participant to move forward with other questions.

For step 3 of the study, we have used the same Google Forms to ask participants about their experience with the chatbot with which they have just interacted. The questions about experience with the Chatbot follow a Likert scale. The open-ended questions and the rationale behind them are listed next.

\begin{itemize}
\item Do you think that chatbot was helpful for the questions you asked? (0 = Helped zero tasks - 5 = Helped in all tasks): We aim to investigate the perception of participants regarding the usefulness of the questions asked during the study.
\item Do you think the chatbot was useful for the presented scenario? (0 = Not useful - 5 = Very useful): We aim to investigate the perception of participants regarding the usefulness of the simple answers planned to be answered during the study.
\item Speed of answer perception? (0= Very slow - 5 = Very fast): We aim to investigate how the users perceive the speed of response of the bot. Because it was a WOZ experiment, participants might perceive the bot as being slow.
\item What other steps do you think this chatbot could cover? (Open question): We aim to investigate the participant’s suggestions of topics that can be covered. 
\item What did you like about the chatbot? (Open question): We aim to investigate positive aspects of the chatbot.
\item What can be improved in the chatbot? (Open question): We aim to investigate constructive criticism by participants.
\item Do you think this solution adds value? Why / why not? (Open question): We aim to investigate if participants think the idea to have software development supported by a smart chatbot is interesting and worthwhile.
\item Do you have any general comments? (Open question):  We aim to investigate if participants have other general comments regarding the interaction with the chatbot.
 \end{itemize}

None of the participants should know which questions the bot is trained to answer, to guarantee they use their vocabulary when interacting with the bot. However, when during the study, participants asked what the bot could do, or when they were asking many questions the bot did not know how to answer, they were given a list of possibilities based on the current  scenario. Every participant was presented with the same scenario. The next section describes the study results.

\section{Study Results}  \label{pilot1results}

\subsection{Step 1: Demographics Collection}

Five students participated in the first part of the study, four masters’ students and one undergraduate. Participants are either from the Computer Science or Software Engineering programs. Regarding their experience with software development, only one participant reported having less than one year experience, as three participants have between one and five years, and one participant had more than five years of experience. Participants have either some knowledge with software repositories (60\%) or good knowledge (40\%).

\subsection{Step 2: Interacting with WOZ Chatbot}

After being presented with the scenario and completing the first part of the questionnaire, participants were supposed to interact with the chatbot. The answers were provided according to keywords in the participants’ questions, and if no expected keyword was present, a default response (Sorry, I can’t help you with that. Is there anything else I can help with?) was presented. Participants interacted with the chatbot, providing an average of 15 interactions/questions. Based on the analysis of the questions, we can highlight the following expected interactions from each participant in Table \ref{tab:expected}. The "x" in the table represents that this participant, in particular, has asked questions regarding what the experiment was ready to answer. The table also shows that all participants asked unexpected questions.

\begin{table}[]
\centering
\caption{Mapping of expected interactions.}
\label{tab:expected}
\begin{tabular}{|l|l|l|l|l|l|}
\hline
\textbf{Expected Intention}  & \textbf{P1} & \textbf{P2} & \textbf{P3} & \textbf{P4} & \textbf{P5} \\ \hline
Greet               & x  & x  & x  & x  & x  \\ \hline
Goodbye             & x  & x  & x  & x  & x  \\ \hline
Artifacts           & x  & x  &    & x  & x  \\ \hline
Commit              & x  & x  &    & x  & x  \\ \hline
Priority            & x  & x  & x  & x  & x  \\ \hline
Pull Request        &    & x  &    & x  & x  \\ \hline
Code Push           &    &    &    &    &    \\ \hline
Sync                &    & x  &    &    & x  \\ \hline
Tasks               & x  & x  & x  & x  & x  \\ \hline
Tool                & x  &    & x  &    &    \\ \hline
Others/fallback     & x  & x  & x  & x  & x  \\ \hline
What can the bot do & x  & x  &    &    & x  \\ \hline
\end{tabular}
\end{table}

\subsection{Semantic Analysis}

We have also attempted to extract entities through semantic analysis (KH Coder tool) of participants' questions. We have loaded stopwords from \url{https://gist.github.com/sebleier/554280}. For word frequency, results of the top 10 words are presented in Table \ref{tab:wordfrequency}.

\begin{table}[]
\centering
\caption{Word Frequency in Participants' Questions, using KH Coder.}
\label{tab:wordfrequency}
\begin{tabular}{|l|l|l|l|l|}
\cline{1-2} \cline{4-5}
\textbf{Word} & \textbf{Frequency} &  & \textbf{Word} & \textbf{Frequency} \\ \cline{1-2} \cline{4-5} 
task          & 20                 &  & need          & 7                  \\ \cline{1-2} \cline{4-5} 
pull          & 9                  &  & artifact      & 5                  \\ \cline{1-2} \cline{4-5} 
request       & 9                  &  & code          & 5                  \\ \cline{1-2} \cline{4-5} 
DevBot        & 8                  &  & commit        & 5                  \\ \cline{1-2} \cline{4-5} 
help          & 8                  &  & create        & 5                  \\ \cline{1-2} \cline{4-5} 
\end{tabular}
\end{table}

We have also analysed the question in clusters, using Jaccard distance, Ward method, and selecting five clusters. The clusters created relate to the questions of: task (18 documents), pull request (9 documents), code (9 documents), greeting (3 documents) and others (40 documents). Nine documents weren’t included in any cluster.

\subsection{Step 3: Questionnaire}

\begin{figure}[]
 \centering
 \includegraphics[scale=0.45]{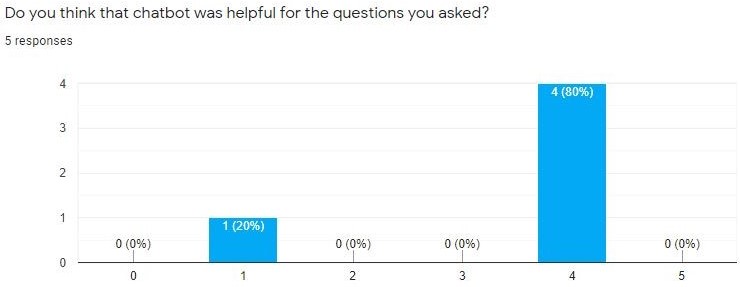}
 \caption{Step 3 - Question 1.}
 \label{fig:1}
\end{figure}

\begin{figure}[]
 \centering
 \includegraphics[scale=0.45]{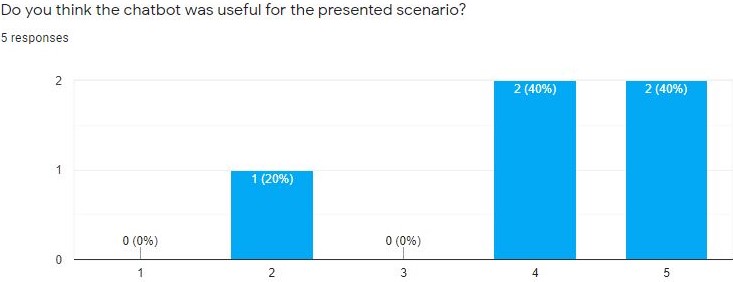}
 \caption{Step 3 - Question 2.}
 \label{fig:2}
\end{figure}

\begin{figure}[]
 \centering
 \includegraphics[scale=0.45]{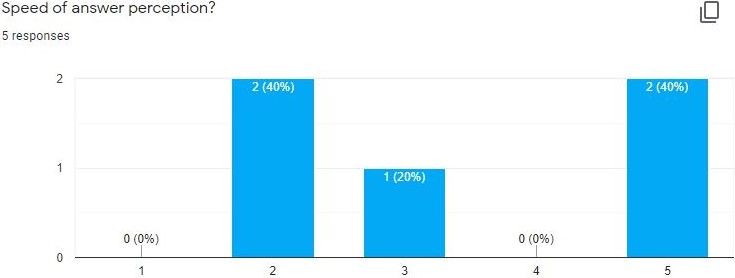}
 \caption{Step 3 - Question 3.}
 \label{fig:3}
\end{figure}

After talking to the bot, participants were asked a few questions regarding the experience with the bot. The results of the Likert-scale questions are presented in Figures \ref{fig:1}, \ref{fig:2} and \ref{fig:3}. 

As of the open questions, the results are presented next.

\textbf{What other steps not covered you think this chatbot could be useful for? }
Respondents have suggested including best practices for coding tips, telling the developer what each task is, providing steps on how to perform tasks, rather than doing it for the developer. Also, respondents have suggested that the chatbot could also help finding out similar tasks to a given particular task, helping to find the experts in some type of task, comparing versions of a file across multiple past commits. Participants also mentioned the need for the chatbot to leverage some context, such as knowing the current repository and pull request comment templates.

\textbf{What did you like about the chatbot?}
Participants suggested that the chatbot was able to clarify what it can be used for, that it was \textit{"helpful in terms of finding my schedule and how my day looks like so I can plan ahead"}. Also another participant has mentioned DevBot \textit{"has the potential to be a good resource for someone who is not very familiar with software development workflows"}.

\textbf{What can be improved in the chatbot?}
In terms of what can be improved, participants have suggested many different aspects:

\begin{itemize}
\item Handling more questions
\item Clarifying its purpose at first
\item Analyze and debug code
\item Offer best practices when coding
\item Know more details regarding what the developer is working on
\item Looking for similar tasks
\item Incorporate webpages (crowd knowledge support)
\item Being able to clarify steps or suggest alternatives instead of stating its inability to help
\end{itemize}

One participant has given the following rich suggestion: \textit{"The leading prompts were helpful but it would have been better if the DevBot could lead with the tasks it is able to complete and how a user can interact with it. Perhaps it could have different modes depending on what git tools, IDEs a user is working with (ie. Eclipse, VScode, vim XD) that could give specific advice for each type of software. A user could select the level of help they require, ie. if a user is brand new, maybe they'd prefer lots of prompts and extra resources for reading about how version control works etc. for an advanced user, maybe they'd prefer just the tasks and a reminder of the workflow..."}

\textbf{Do you think this solution adds value? Why / why not?} When asked if the chatbot adds value, most participants answered that it does adds value and that a solution such as DevBot can be really useful for software developers.

\textbf{Do you have any general comments?} Respondents have raised that it would be interesting to have the chatbot integrated into a software development workflow as a personal assistant and adding personality to the chatbot.

\subsection{Discussion}

Most of the questions asked by participants were in the subject of tasks and how to implement development steps. Therefore, there are indications that developers are interested in being guided in their tasks rather than having a chatbot that can execute tasks on their behalf. We find this interesting, as managing tasks is not essential to contributing code per se but is perceived to be important, hence there might be this gap in expectations between CA designers and users that must be taken into account.

Most respondents have suggested that adding context to the chatbot would be interesting. For example, the chatbot should know the repository address, branch or patterns expected to commit messages. However, the simple study with a few expected questions and given a simple scenario was also indicated as being helpful by most of the participants. Based on feedback from participants, adding context to the chatbot could immensely improve the chatbot’s capabilities, as it would be personalized. The chatbot could also incorporate discovery of context or recommendations based on the history of interactions.

Participant P1, the least experienced developer (less than 1 year according to demographics), was the participant that  indicated a low score when asked if the chatbot was helpful for the questions asked.  This can indicate that the chatbot has to be prepared to give more details about the process to less experienced users. All other participants, who have more than one year of experience, indicated that the chatbot was helpful.

Two participants have asked questions regarding the tool, interestingly, the ones that have not done questions regarding pull requests. We believe that this can indicate less experience developing software and interacting with Git as version control repositories. When analyzing if participants took time to create the questions, the perception was that the questions were created quickly, as the study had an average of 15 questions asked in an average period of 10.8 minutes of chatting with the bot actively.

All participants have greeted the bot at the beginning of the interaction and when leaving the chat. Only one participant realized it was a WOZ experiment. This demonstrates that the bot should be able to handle greeting interactions. None of the participants asked questions regarding pushing code to their local repository or pulling, merging or cloning. This could indicate developers are not worried about these tasks, that are done less often during development. However, most participants asked about pull requests (requesting the master repository’s owner to pull files from your repository), which can happen quite often during software development.


\section{Conclusion} \label{conclusion}

We have presented an exploratory study regarding the preferences of software developers when interacting with a chatbot to support work. Developers usually work with various tools in a very dynamic situation, and promoting ways to support them is vital for the quality of their work. As well, a chatbot could improve productivity, lessen training times and make implicit preferences explicit through capture processes. Our results demonstrate developers are willing to work with such tools and have found this solution to be interesting, providing ideas for the future and hints on how chatbots could behave and connect.

As future work, we intend to implement a real chatbot using Rasa and run this study with more participants, while also developing a context model that is capable of supporting the interactions with developers, storing their work variables (context) as indicated by the participants of this pilot study and the literature. We believe such types of solutions can be useful for teams that work in shared office space and especially for teams that work remotely.

\section*{Acknowledgements}

The authors thank the Natural Sciences and Engineering Research Council of Canada (NSERC), the MITACS Accelerate Program, and the Centre for Community Mapping (COMAP).

\bibliographystyle{IEEEtran} 
\bibliography{main}

\begin{thebibliography}{10}
\providecommand{\url}[1]{#1}
\csname url@samestyle\endcsname
\providecommand{\newblock}{\relax}
\providecommand{\bibinfo}[2]{#2}
\providecommand{\BIBentrySTDinterwordspacing}{\spaceskip=0pt\relax}
\providecommand{\BIBentryALTinterwordstretchfactor}{4}
\providecommand{\BIBentryALTinterwordspacing}{\spaceskip=\fontdimen2\font plus
\BIBentryALTinterwordstretchfactor\fontdimen3\font minus
  \fontdimen4\font\relax}
\providecommand{\BIBforeignlanguage}[2]{{%
\expandafter\ifx\csname l@#1\endcsname\relax
\typeout{** WARNING: IEEEtran.bst: No hyphenation pattern has been}%
\typeout{** loaded for the language `#1'. Using the pattern for}%
\typeout{** the default language instead.}%
\else
\language=\csname l@#1\endcsname
\fi
#2}}
\providecommand{\BIBdecl}{\relax}
\BIBdecl

\bibitem{bradley2018context}
N.~Bradley, T.~Fritz, and R.~Holmes, ``Context-aware conversational developer
  assistants,'' in \emph{2018 IEEE ACM 40th International Conference on
  Software Engineering (ICSE)}.\hskip 1em plus 0.5em minus 0.4em\relax IEEE,
  2018, pp. 993--1003.

\bibitem{meyer2017work}
A.~N. Meyer, L.~E. Barton, G.~C. Murphy, T.~Zimmermann, and T.~Fritz, ``The
  work life of developers: Activities, switches and perceived productivity,''
  \emph{IEEE Transactions on Software Engineering}, vol.~43, no.~12, pp.
  1178--1193, 2017.

\bibitem{melo2019retrieving}
G.~Melo, T.~Oliveira, P.~Alencar, and D.~Cowan, ``Retrieving curated stack
  overflow posts from project task similarities,'' in \emph{International
  Conference on Software Engineering Knowledge Engineering}, 2019, pp.
  415--418.

\bibitem{salber1999context}
D.~Salber, A.~K. Dey, and G.~D. Abowd, ``The context toolkit: aiding the
  development of context-enabled applications,'' in \emph{Proceedings of the
  SIGCHI conference on Human Factors in Computing Systems}, 1999, pp. 434--441.

\bibitem{murphy2019beyond}
G.~C. Murphy, ``Beyond integrated development environments: adding context to
  software development,'' in \emph{2019 IEEE/ACM 41st International Conference
  on Software Engineering: New Ideas and Emerging Results (ICSE-NIER)}.\hskip
  1em plus 0.5em minus 0.4em\relax IEEE, 2019, pp. 73--76.

\bibitem{gasparic2017context}
M.~Gasparic, G.~C. Murphy, and F.~Ricci, ``A context model for ide-based
  recommendation systems,'' \emph{Journal of Systems and Software}, vol. 128,
  pp. 200--219, 2017.

\bibitem{melo2019context}
G.~Melo, P.~Alencar, and D.~Cowan, ``Context-augmented software development in
  traditional and big data projects: Literature review and preliminary
  framework,'' in \emph{2019 IEEE International Conference on Big Data (Big
  Data)}.\hskip 1em plus 0.5em minus 0.4em\relax IEEE, 2019, pp. 3449--3457.

\bibitem{marieli2018}
\BIBentryALTinterwordspacing
M.~Wessel, B.~M. de~Souza, I.~Steinmacher, I.~S. Wiese, I.~Polato, A.~P.
  Chaves, and M.~A. Gerosa, ``The power of bots: Characterizing and
  understanding bots in oss projects,'' \emph{Proceedings of the ACM on
  Human-Computer Interaction}, vol.~2, no. CSCW, Nov. 2018. [Online].
  Available: \url{https://doi.org/10.1145/3274451}
\BIBentrySTDinterwordspacing

\bibitem{weizenbaum1966eliza}
J.~Weizenbaum, ``Eliza—a computer program for the study of natural language
  communication between man and machine,'' \emph{Communications of the ACM},
  vol.~9, no.~1, pp. 36--45, 1966.

\bibitem{mctear2016rise}
M.~F. McTear, ``The rise of the conversational interface: A new kid on the
  block?'' in \emph{International Workshop on Future and Emerging Trends in
  Language Technology}.\hskip 1em plus 0.5em minus 0.4em\relax Springer, 2016,
  pp. 38--49.

\bibitem{storey2016disrupting}
M.-A. Storey and A.~Zagalsky, ``Disrupting developer productivity one bot at a
  time,'' in \emph{Proceedings of the 2016 24th ACM SIGSOFT International
  Symposium on Foundations of Software Engineering}, 2016, pp. 928--931.

\bibitem{paikari2019chatbot}
E.~Paikari, J.~Choi, S.~Kim, S.~Baek, M.~Kim, S.~Lee, C.~Han, Y.~Kim, K.~Ahn,
  C.~Cheong \emph{et~al.}, ``A chatbot for conflict detection and resolution,''
  in \emph{2019 IEEE/ACM 1st International Workshop on Bots in Software
  Engineering (BotSE)}.\hskip 1em plus 0.5em minus 0.4em\relax IEEE, 2019, pp.
  29--33.

\bibitem{cerezo2019building}
J.~Cerezo, J.~Kubelka, R.~Robbes, and A.~Bergel, ``Building an expert
  recommender chatbot,'' in \emph{2019 IEEE/ACM 1st International Workshop on
  Bots in Software Engineering (BotSE)}.\hskip 1em plus 0.5em minus 0.4em\relax
  IEEE, 2019, pp. 59--63.

\bibitem{gnewuch2017towards}
U.~Gnewuch, S.~Morana, and A.~Maedche, ``Towards designing cooperative and
  social conversational agents for customer service.'' in \emph{Proceedings of
  the International Conference on Information Systems (ICIS)}, 2017.

\bibitem{lee2009example}
C.~Lee, S.~Jung, S.~Kim, and G.~G. Lee, ``Example-based dialog modeling for
  practical multi-domain dialog system,'' \emph{Speech Communication}, vol.~51,
  no.~5, pp. 466--484, 2009.

\bibitem{radziwill2017evaluating}
N.~M. Radziwill and M.~C. Benton, ``Evaluating quality of chatbots and
  intelligent conversational agents,'' \emph{arXiv preprint arXiv:1704.04579},
  2017.

\bibitem{brenier_prial_2019}
\BIBentryALTinterwordspacing
J.~Brenier and J.~Prial, ``An overview of conversational ai,'' May 2019,
  accessed on = 2020-02-27. [Online]. Available:
  \url{https://georgianpartners.com/investment-thesis-areas/overview-conversational-ai/}
\BIBentrySTDinterwordspacing

\bibitem{bocklisch2017rasa}
T.~Bocklisch, J.~Faulkner, N.~Pawlowski, and A.~Nichol, ``Rasa: Open source
  language understanding and dialogue management,'' \emph{arXiv preprint
  arXiv:1712.05181}, 2017.

\bibitem{singh2019introduction}
A.~Singh, K.~Ramasubramanian, and S.~Shivam, ``Introduction to microsoft bot,
  rasa, and google dialogflow,'' in \emph{Building an Enterprise
  Chatbot}.\hskip 1em plus 0.5em minus 0.4em\relax Springer, 2019, pp.
  281--302.

\bibitem{woz}
\BIBentryALTinterwordspacing
J.~D. Gould, J.~Conti, and T.~Hovanyecz, ``Composing letters with a simulated
  listening typewriter,'' \emph{Commun. ACM}, vol.~26, no.~4, p. 295–308,
  Apr. 1983. [Online]. Available: \url{https://doi.org/10.1145/2163.358100}
\BIBentrySTDinterwordspacing

\bibitem{kelleydid}
\BIBentryALTinterwordspacing
J.~Kelley, ``Where did the usability term wizard of oz come from.'' [Online].
  Available: \url{http://www.musicman.net/oz.html}
\BIBentrySTDinterwordspacing

\end{thebibliography}

\end{document}